\documentclass[aps,prx,reprint,superscriptaddress,nofootinbib,longbibliography,onecolumn,11pt]{revtex4-2}
\usepackage[pdfpagelabels=false,colorlinks=true,allcolors=black]{hyperref}  

\usepackage[export]{adjustbox}
\usepackage{booktabs}
\usepackage{float}
\usepackage{flushend}
\usepackage{graphicx}
\usepackage{grffile}
\usepackage{ifpdf}
\usepackage[utf8]{inputenc}
\usepackage{listings}
\usepackage{multirow}
\usepackage[caption=false,font=footnotesize]{subfig}
\usepackage{url}
\usepackage{tabularx}
\usepackage{textcomp}
\usepackage{xcolor}
\usepackage{xspace}
\usepackage{comment}
\usepackage{outlines}
\usepackage{tikz} 

\usepackage[pagewise, modulo]{lineno}

\definecolor{listinggray}{gray}{0.95}
\definecolor{darkgray}{gray}{0.7}
\definecolor{commentgreen}{rgb}{0, 0.4, 0}
\definecolor{darkblue}{rgb}{0, 0, 0.6}
\definecolor{purple}{rgb}{0.6, 0, 0.6}
\definecolor{middleblue}{rgb}{0, 0, 0.75}
\definecolor{darkred}{rgb}{0.4, 0, 0}
\definecolor{brown}{rgb}{0.5, 0.5, 0}
\definecolor{dkgreen}{rgb}{0,0.5,0}
\definecolor{orange}{rgb}{1,.5,0}
\definecolor{dandelion}{cmyk}{0,0.29,0.84,0}

\usepackage[normalem]{ulem}
\makeatletter
\def\cyanuwave{\bgroup \markoverwith{\lower3.5\p@\hbox{\sixly \textcolor{cyan}{\char58}}}\ULon}
\def\reduwave{\bgroup \markoverwith{\lower3.5\p@\hbox{\sixly \textcolor{red}{\char58}}}\ULon}
\def\blueuwave{\bgroup \markoverwith{\lower3.5\p@\hbox{\sixly \textcolor{blue}{\char58}}}\ULon}
\font\sixly=lasy6 
\makeatother

\def\BibTeX{{\rm B\kern-.05em{\sc i\kern-.025em b}\kern-.08em
    T\kern-.1667em\lower.7ex\hbox{E}\kern-.125emX}}

\newif\ifdraft{}
\drafttrue{}

\ifdraft{}
  \newcommand{\amnote}[1]{ \textcolor{blue} { ***andrem: #1 }}
  \newcommand{\jhanote}[1]{ {\textcolor{red} { ***shantenu: #1 }}}
  \newcommand{\mtnote}[1]{ {\textcolor{orange} { ***matteo: #1 }}}
  \newcommand{\vrpnote}[1]{ {\textcolor{teal} { ***vrp: #1 }}}
  \newcommand{\fixme}{{\textcolor{red}{FIXME \xspace}}}

\else
  \newcommand{\amnote}[1]{}
  \newcommand{\jhanote}[1]{}
  \newcommand{\mtnote}[1]{}
  \newcommand{\vrpnote}[1]{}
  \newcommand{\fixme}[1]{}
\fi

\lstdefinestyle{myListing}{
  frame=single,
  backgroundcolor=\color{listinggray},
  language=C,
  basicstyle=\ttfamily \footnotesize,
  breakautoindent=true,
  breaklines=true
  tabsize=2,
  captionpos=b,
  aboveskip=0em,
  belowskip=-2em,
}

\lstdefinestyle{myPythonListing}{
  frame=single,
  backgroundcolor=\color{listinggray},
  language=Python,
  basicstyle=\ttfamily \footnotesize,
  breakautoindent=true,
  breaklines=true
  tabsize=2,
  captionpos=b,
}

\ifpdf{}
  \DeclareGraphicsExtensions{.pdf, .jpg, .tif}
\else
  \DeclareGraphicsExtensions{.eps, .jpg, .ps}
\fi

\begin{document}
\title{AI-coupled HPC Workflows}

\author{Shantenu Jha}
\email{shantenu@bnl.gov}
\affiliation{Rutgers University, New Brunswick, NJ 08901}
\affiliation{Computational Science Initiative, Brookhaven National Laboratory, Upton, NY 11972, USA}
\author{Vincent R. Pascuzzi}
\email{pascuzzi@bnl.gov}
\affiliation{Computational Science Initiative, Brookhaven National Laboratory, Upton, NY 11972, USA}
\author{Matteo Turilli}
\email{mturilli@bnl.gov}
\affiliation{Rutgers University, New Brunswick, NJ 08901}
\affiliation{Computational Science Initiative, Brookhaven National Laboratory, Upton, NY 11972, USA}

\begin{abstract}
Increasingly, scientific discovery requires sophisticated and scalable
workflows. Workflows have become the ``new applications,'' wherein multi-scale
computing campaigns comprise multiple and heterogeneous executable tasks. In
particular, the introduction of AI/ML models into the traditional HPC
workflows has been an enabler of highly accurate modeling, typically reducing
computational needs compared to traditional methods. This chapter discusses
various modes of integrating AI/ML models to HPC computations, resulting in
diverse types of AI-coupled HPC workflows. The increasing need of coupling
AI/ML and HPC across scientific domains is motivated, and then exemplified by
a number of production-grade use cases for each mode. We additionally discuss
the primary challenges of extreme-scale AI-coupled HPC campaigns---task
heterogeneity, adaptivity, performance---and several
framework and middleware solutions which aim to address them. While both HPC
workflow and AI/ML computing paradigms are independently effective, we
highlight how their integration, and ultimate convergence, is leading to
significant improvements in scientific performance across a range of domains,
ultimately resulting in scientific explorations otherwise unattainable.
\end{abstract}

\date{\today}
\maketitle



\setcounter{page}{1}

\section{Introduction}\label{chr-intro}


Scientific discovery increasingly requires sophisticated and scalable
workflows. Workflows have become the ``new applications,'' wherein multi-scale
computing campaigns comprise hundreds to thousands of heterogeneous executable
tasks. Introducing AI/ML models into traditional high performance computing
(HPC) workflows has been an enabler of highly accurate modeling, and has been
demonstrated to be a promising approach for significant performance
improvements.

Advances in statistical algorithms and runtime systems have enabled extreme
scale ensemble-based applications~\cite{cosb18kasson} to overcome limitations
of traditional monolithic simulations. However, in spite of several orders of
magnitude improvement in efficiency from these ensemble algorithms, the
complexity of phase space and dynamics for modest physical systems require
additional orders of magnitude improvements and performance gains. Integration
of traditional HPC workflows with AI/ML methods holds real promise for
overcoming such barriers~\cite{hruska2019extensible}.






In many application domains, the integration of AI/ML into a computational
workflow is a favorable way to obtain large performance gains, and presents an
opportunity to jump a generation of simulation enhancements. For example, one
can view the use of learned surrogates as a performance boost that can lead to
substantial speedups, as calculation of a prediction from a trained network can
be many orders of magnitude faster than full execution of the
simulation~\cite{Fox2019Learning,Kasim20published}. In addition to the use of
learning for advanced sampling as mentioned above, other simple examples
include the use of a surrogate to represent a chemistry
potential~\cite{9355242}, or a larger grain size to solve the diffusion
equation underlying cellular and tissue level
simulations~\cite{peterson2019merlin}.

There are various modes (couplings) of integrating traditional HPC methods and
simulations, with AI/ML methodologies, resulting in diverse types of AI/ML
``enhanced'' HPC workflows. This chapter provides an overview the various
couplings and how they can result in the adaptive execution of workflow
applications comprising heterogeneous tasks. We identify the core
characteristics of such workflow applications, as well as discuss
state-of-art tools and workflow applications.



\section{Learning EveryWhere Paradigm}\label{chr-lep}


There are two classes of interplay between HPC and ML. In the first, ML
directly enhances and impacts applications; in the second class, ML enhances
the HPC environment on which those applications operate. This chapter
exclusively focuses on the former.

Central to the first class, as well as the re-examination and overcoming the
performance barrier, is the need to integrate ML methodologies and HPC. In
this approach \textendash{} learning enhanced simulations and campaigns
\textendash{} we include the use of neural surrogates, with a neural network
directly predicting either the full results of simulations, or components
thereof. This also includes using learning methods to control and steer
simulations, for example, efficient campaigns that steer ensembles smartly
through phase
space\cite{lee2019deepdrivemd,lee2020scalable-short,saadi2020impeccable-short,casalino2020aidriven-short}.
We have identified three high-level modes of integrating ML with
HPC~\cite{Fox2019Learning,jha2019understanding,fox2019taxonomy}:
\textbf{ML-in-HPC}, \textbf{ML-out-HPC}, and \textbf{ML-about-HPC}.

\begin{figure}
  \centering
    \includegraphics[width=0.8\textwidth]{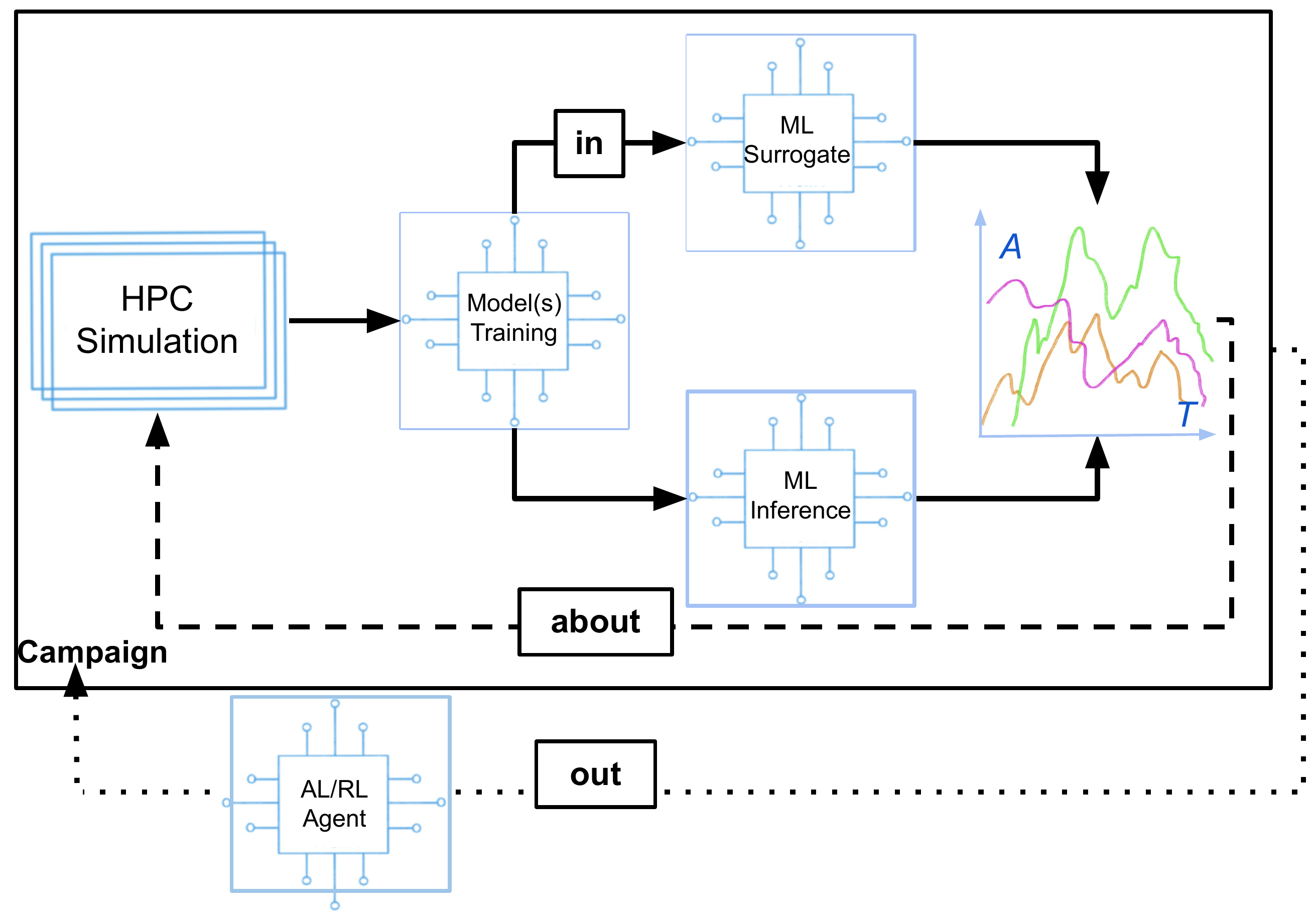}
    \caption{Illustrating the three modes of ML-x-HPC: (1) ML-in-HPC: AI/ML
surrogate models are used to replace part or entire simulations.  (2)
ML-about-HPC: AI/ML complements traditional computational tasks and possibly steers the tasks, improving
their scientific results or efficiency (3) ML-out-HPC: A high-level AI/ML based
algorithm, such as active learning or reinforcement learning is used to
dynamically control the campaign, or steer the workflow as a whole (as
opposed to just the tasks). Typically, ML-in and ML-about are directly
responsible for producing output for further analysis, while ML-out drives this production.}
  \label{fig:ml-modes}
\end{figure}

\textbf{ML-in-HPC} represents the scenario when an ML model is introduced in
lieu of a component of the HPC simulations, or possibly, in lieu of the total
simulation itself, i.e., ML model serves as a ``total surrogate''.
\textbf{ML-out-HPC} captures situations wherein a ML model resides ``outside''
of the traditional HPC simulation loop, but dynamically controls the
progression of the HPC workflow. For example, Active Learning and
Reinforcement Learning control of computational campaigns. Finally,
\textbf{ML-about-HPC} represents the situation where ML models are concurrent
and coupled to the main HPC tasks. Figure~\ref{fig:ml-modes} illustrates
these primary modes coupling AI/ML to HPC workflows. These three modes are not
mutually exclusive, and will increasingly be used collectively.

The \textit{learning everywhere}
paradigm~\cite{Fox2019Learning,fox2019taxonomy,jha2019understanding} contends
that increasingly, scientific applications will achieve performance gains and
methodological advances by  using all three modes of combining learning
approaches with HPC simulation-based techniques. In the next section, we will
discuss multilevel drug selection as a canonical example of
learning-everywhere paradigm, but additional prominent examples include
materials design and earth-systems modeling~\cite{Kasim20published}.



There are many open challenges that implementing and translating the paradigm
to practice poses. For example, how and where can ML effectively enhance or
accelerate HPC simulations? How to make ML methods that work in tandem with
HPC simulations scalable, robust, and reliable? For a given computational
campaign what is the optimal mix and execution plan of ML-in, out and about
HPC? Furthermore, there are system and software challenges and opportunities
in combining ML and HPC systems software, hardware, and overall
infrastructure. What are the correct programming models and abstractions to
manage the diverse ``computational tasks'' viz., ML training \& inference
along with traditional HPC workloads? What runtime systems are needed to
manage the heterogeneous workload effectively? What are the general motifs of
interaction between ML and HPC, and their influence on design of runtime
systems?



A leitmotif of the learning everywhere paradigm is \textbf{effective
performance}, i.e., the performance improvement obtained by substituting a
traditional HPC method with an integrated HPC and learning method. Effective
performance measures the improvement in application performance metric (e.g.,
computational cost, improved time-to-solution, or the achievement of
scientific objective) when using ML methods in conjunction with HPC, as
compared to using HPC methods stand-alone.
For example, effective performance can be measured as
the time-to-solution ratio of the traditional approach vs.\ the
learning-enhanced approach. If a traditional parameter study ran 1000
simulations to determine an optimal engineering design, while a model-based
optimizer produced the same optimum in 100 simulations, the effective
performance of the learning enhanced application is 10. If, additionally, the
ML-based approximations in the simulation accelerated computation by a factor
of 10, the effective performance would be 100. These orders-of-magnitude
increase in effective performance as learning-enhanced high-performance
computing takes root~\cite{RN89} are at the heart of the motivation for this
new paradigm of computation.




\section{Learning EveryWhere Examples}\label{chr-lee}

This section provides an overview of various use cases and exemplar
applications, across scientific domains which couple HPC and ML/AI.
Table~\ref{tab:mlhpc-uc} groups use cases and exemplar applications using the
three modes described in Sec.~\ref{chr-lep}. Use cases and
applications were selected to provide a representative overview of the ML techniques currently
employed to couple ML with HPC, and to cover diverse scientific domains in
which this coupling is bringing innovation and unprecedented performance
improvement.

\begin{table*}
  \centering
  \resizebox{\textwidth}{!}{
    \begin{tabular}{llll}
      \toprule
      \textbf{Mode}                           &
      \textbf{Domain}                         &
      \textbf{Application}                    &
      \textbf{Coupling Mechanism}             \\
      \toprule
      \multirow{2}{*}{ML-in-HPC}              &
      High Energy physics                     &
      Atlfast3                                &
      Surrogate methods                       \\
                                              &
      Molecular Dynamics                      &
      DeePMD-kit                              &
      Surrogate methods                       \\
      \midrule
      \multirow{3}{*}{ML-about-HPC}           &
      Atomistic simulations                   &
      Proxima                                 &
      Runtime surrogate tuning                \\
                                              &
      Material engineering                    &
      Colmena                                 &
      Runtime model (re)training/configuration \\
                                              &
      RAS protein/Cancer                      &
      MuMMI                                   &
      Runtime ML-based selection              \\
      \midrule
      \multirow{3}{*}{ML-out-HPC}             &
      Cancer research                         &
      DeepHyper                               &
      Automated machine learning              \\
                                              &
      Cyberinfrastructure                     &
      SIOX                                    &
      Offline ML                              \\
                                              &
      Materials Science                       &
      EXARL                                   &
      ML-guided simulations                   \\
      \bottomrule
    \end{tabular}
  }  \caption{Examples of ML-x-HPC modes across scientific domains.}
\label{tab:mlhpc-uc}
\end{table*}


%

\subsection{ML-in-HPC}

Workflows in high energy physics are multi-scale, comprising quantum field
theoretic calculations, detector simulations, and classical reconstruction
of physical objects. Each scale has considerable computational requirements and,
using only traditional methods, it is impossible to produce sufficient numbers
of Monte Carlo events to maintain statistical adequacy with recorded data.
%
As such, ML techniques, including generative adversarial networks, are
becoming an increasingly attractive alternative to standard frameworks
implementing step-by-step model predictions.

\begin{figure}
  \centering
    \includegraphics[width=0.8\textwidth]{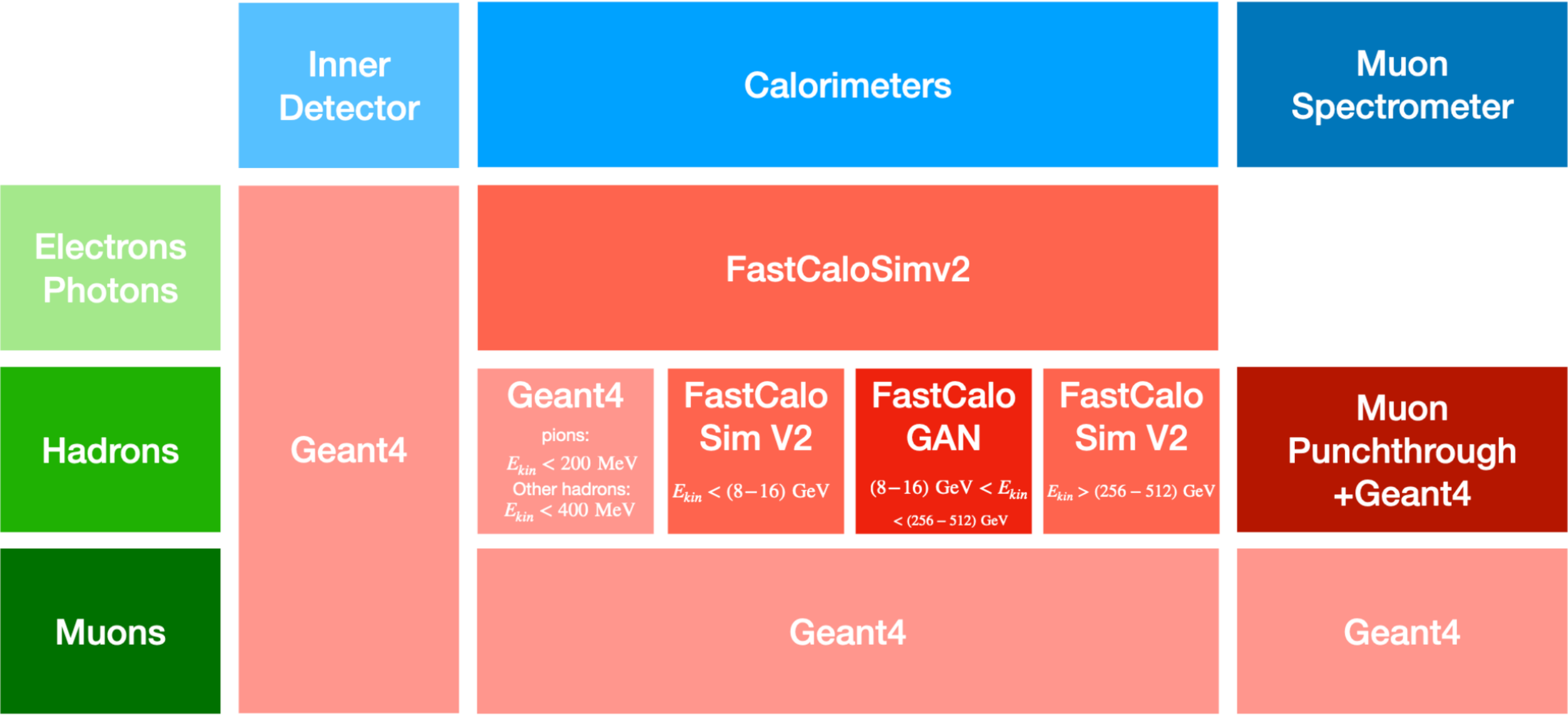}
    \caption{Illustration of ML-in-HPC, showing a configuration of different
    surrogate modules replacing Geant4 in the ATLAS detector. Different
    ML-based components can be employed, depending on the type of process being
    modeled, specific subdetectors, particle types and energies.
    Image from~\cite{atlfast3}.}
  \label{fig:atlfast3}
\end{figure}

For example, Fig.~\ref{fig:atlfast3} shows an illustration of the ATLAS
Collaboration's ``fast'' simulation framework, Atlfast3~\cite{atlfast3}, where
ML-based techniques are used in place of intensive Geant4~\cite{geant4}
simulations. In this configuration, surrogates are employed in lieu of full
Geant4 simulations for specific particle types, energies and subdetectors to
reduce overall simulation times up to several orders of magnitude.
At the same time, those surrogates allow to maintain accurate detector
modeling for new physics searches, statistically-limited analyses, background
processes, and detector upgrades.
%



Another significant example of ML-in-HPC can be found in the traditional
\textit{ab initio} MD (AIMD)
methods for modeling atomistic phenomena. Due to demanding computational
requirements (cubic scaling in the number of electronic degrees of freedom),
most AIMD applications are limited to $O(1000)$ atoms. However, AIMD plays a
major role in addressing many issues related to, e.g., drug discovery,
complex chemical processes and nanotechnology. As such, tremendous efforts have
been afforded to more efficient methods, including ML.

Jia \textit{et al.}~\cite{9355242} offer a powerful example of ML-in-HPC
applied to AIMD.
Jia's approach employs an ML-based simulation protocol which uses surrogates
(Deep Potential MD) in conjunction with a highly-optimized code (a
GPU-accelerated DeePMD-kit~\cite{WANG2018178}) to simulate $O(10^8)$
nanosecond-long trajectories in 24 hours. This record-setting accomplishment
efficiently scaled to the whole 4,560 nodes of the Summit supercomputer,
reaching double/mixed-single/mixed-half precision performance of 91/162/275
PFLOPS. Compared to other state-of-the-art, Jia \textit{et al.} showed more than
$O(10^{-3})$ reduction in time-to-solution (TTS) [s/step/atom].

Ref.~\cite{9355242} is a prime example of how ML-in-HPC, which is on the cusp
of a paradigmatic change as learning approaches influence the way both ODE and
PDEs are being solved. For example, Karniadakis et al~\cite{raissi2019physics}
are investigating how to solve and discover new PDEs via deep learning. For
partial differential equations (PDEs), neural operators directly learn the
mapping from any functional parametric dependence to the solution. Thus, they
learn an entire family of PDEs, in contrast to classical methods which solve
one instance of the equation~\cite{li2020fourier}.


\subsection{ML-about-HPC}


%

Replacing computationally intensive computations with surrogate approximators
aims to reduce TTS often by sacrificing accuracy with respect to more complete
models.
Optimally balancing TTS and accuracy is a non-trivial task, and
the former is often neglected in order to reach the desired accuracy.
Proxima~\cite{proxima}, provides real-time feedback from executing simulations
and it has been utilized to develop systematic and automated methods for
dynamically tuning surrogate configurations. An iterative simulation workflow,
representative of the ML-about-HPC mode, evaluates uncertainties
associated with the use of surrogates, concurrently updating configurations based
on a distance metric to learn features and an accuracy metric to evaluate
prediction. Between iterations, surrogates evolve and replace less optimized
ones, providing the coupling between concurrent surrogate tuning and
the main HPC campaign.

The Proxima framework has been demonstrated in a Mone Carlo sampling
application, where the first-principles Hartree-Fock~\cite{osti_5156729}
prediction target is replaced with a Proxima-managed surrogate. Mean absolute
error (MAE) and TTS comparisons are made between Proxima and a surrogate
strategy with a fixed distance threshold (based on scientific trial-and-error).
In certain scenarios, the fixed strategy outperforms Proxima in terms of TTS.
However, the utility of user-defined error bounds ensures more robust results
with Proxima. By determining values for surrogate configurations automatically
during workflow execution, Proxima is able to satisfy error bounds while
achieving as much as 5.5x speedup in TTS.

Estimating properties of large collections of molecules is often necessary to
find candidates for medical therapeutics, next-generation batteries, etc.
However, the number of possible candidates for a single application, and
therefore the number of different experimental configurations required to test
them all, is often intractable. Large-scale workflows have thus come to adopt
methodologies to provide an ML model training and retraining runtime to decide
which computations to perform based on previous outputs.

The Colmena framework~\cite{ward2021colmena} facilitates a
user-defined steering for workflow execution.
Using an example application involving an ML-guided search of $10^5$ molecules
with high resistance to oxidation electrolyte design, the Colmena workflow
provides components to actively train and learn as simulations are executed.
Colmena performs a concurrent execution mode, having ML and traditional
simulations running side-by-side throughout the workflow.
Candidacy of molecules is evaluated using ML models which are scored based on
selection criteria and then are ordered based on their score. Molecules
appearing at the top of the ordered list reflect most suitable candidates, and
subsequently additional simulations are executed. 
Colmena is reported to find
candidate molecules at rates 100 times that of traditional computational solutions, and scaling up to 1024 nodes (65,636 cores) on Theta
supercomputer.




A naturally more complicated scenario is the development of therapeutics. This
type of R\&D can take years or decades to come to fruition due to the
complicated computational modeling involved in searches for candidate drugs and
strict FDA approval procedure. In the context of cancer treatments, for example,
it is suggested that Ras proteins are involved in nearly a third of all human
cancers in the US~\cite{SIMANSHU201717}; however, many physiochemical properties of
Ras-Raf-membrane dynamics are not fully understood. The inherent multi-scale
nature of such processes makes computational modeling challenging, and each
scale is traditionally simulated separately.

The massively parallel Multiscale Machine-Learned Modeling Infrastructure
(MuMMI)~\cite{mummi1-short,mummi2-short}
couples three resolution scales with ML-based selection, effectively promoting
important configurations from coarse-grained to all atomistic (highest
resolution). The autonomy and full power of MuMMI is realized through dynamic
co-scheduling of tasks which is achieved by tying together application and
coordination layers of the workflow. MuMMI achieves a
98\% GPU occupancy for more than 83\% of 600,000 node hours, coordinating 24,000
jobs and managing several terabytes of data daily. Furthermore, the split
architecture, separating the workflow application from coordination, permits
generalizability, making the infrastructure attractive beyond drug design.


\subsection{ML-out-HPC}

Increasing computational power helps to produce more rapidly predictive models,
larger volumes of collected data enables higher fidelity predictions. However,
improving models typically implies introducing additional complexity such as
substantially increasing trainable parameters. As such, building ML models for
complex diseases---such as cancer---involves a significant amount of
trial-and-error, and intervention from both epidemiologists and ML experts,
making diagnosis, detection, prognosis and prediction extremely time-consuming
tasks.

Work from Balaprakash \textit{et al.}~\cite{balaprakash2019scalable}
introduces a reinforcement, learning-based neural architecture search for
autonomous deep learning development. By targeting specific class of cancer
data, the automated approach finds neural architectures requiring fewer
trainable parameters---thus reducing training time---which produce equivalent or better
accuracy to manually finely-tuned architectures. Scalability is demonstrated
using 1024 nodes of the Theta supercomputer, with the best neural architecture
outperforming the manually designed network in terms of scientific results, and
having $11.5\times$ fewer trainable parameters and $2.5\times$ faster training
time. These results suggest ML-driven neural architecture search has the
potential to accelerate cancer research, allowing researchers in the field to
automate neural architecture discovery using HPC.

To accommodate needs of scientists and non-ML experts, a recent effort
providing flexible user tools for distributed and scalable reinforcement
learning (RL) is the EXARL~\cite{exarl-github} from the Co-Design Center for
Exascale Machine Learning Technologies (ExaLearn).
EXARL enables interfacing with existing exascale applications---\textit{e.g.},
LAMMPS~\cite{LAMMPS} and NWChem~\cite{nwchem}---which domain scientists
can guide using RL algorithms and associated neural network architectures.
In addition to the miniGAN proxy application~\cite{exarl}, the ExaLearn team
has demonstrated its usefulness and exercise its scalable RL to a block
copolymer application~\cite{doi:10.1177/10943420211029302}.
This is generally a complicated problem as materials may evolve
toward generic states or become trapped in a metastable state, and thus
requiring hundreds of experimental trials to reach a target state.
By mapping this problem to RL, wherein a NN is trained to update annealing
temperatures for subseqeunt block copolymer simulations, EXARL was able to show
learning convergence for guiding the annealing process to both equilibrium
and non-equilibrium states.
Ongoing and future work includes expanding to new scientific domain use-cases,
and enablement of further scaling and execution of multi-process applications.



While HPC drives much of scientific research and discovery, the platforms
themselves require continuous performance analysis and optimizations to reach
their full potential. This is particularly difficult for I/O systems which are
commonly bottlenecks in computing systems, trailing in performance with respect
to computational capabilities by several orders of magnitude. This is due to
complexity of I/O systems, requiring intimate knowledge of the underlying
components and potentially thousands of parameters need to mutual optimization.

The SIOX Project~\cite{kunkel2014siox}
monitors, diagnoses and optimizes I/O system parameters of HPC platforms.
The modular design of SIOX provides plug-and-play capability, allowing to use
diverse monitoring tools for data production. Plugins use offline ML to predict
the performance gains or losses for different optimization actions
and online ML to perform anomaly detection.
SIOX implements actuator tasks to apply the selected optimizations and evaluator
tasks to measure achieved performance.

\subsection{Learning EveryWhere: A Canonical Example}

The three modes of coupling ML with HPC are not mutually exclusive. In fact,
the most ambitious multi-scale or multi-stage campaigns involve all three
modes. For example, considering the universe of about 10$^{68}$ possible drug
compounds, efficient and high throughput frameworks for early stage drug
discovery~\cite{Bohacek_et_al:2010} are needed. \textit{In silico}
methodologies need to be improved to better select lead compounds that can
proceed to later stages of the drug discovery protocol accelerating the entire
process~\cite{antunes2015understanding,zhou2020artificial,smith2018transforming}.
Innovations that integrate AI and simulation at multiple levels are
demonstrating promise in overcoming fundamental limitations.

We discuss IMPECCABLE as a representative campaign that is comprised of
ML-in-, ML-out-, and ML-about-HPC workflows, supplanting traditional HPC with
learning everywhere. Although, IMPECCABLE was developed for COVID19
therapeutics, the multi-stage and AI-HPC integrated campaign is representative
of a range of campaigns in material and molecule design, climate science,
inter alia.

The campaign consists of an iterative loop initiated with ML predictions
(ML1), followed by data processing stages S1, S2, S3. ML/AI techniques (ML1
and S2) interfaced with physics-based methods estimate docking poses of
compounds that are promising leads for a given protein target (S1) and binding
free-energy computations (S3). Put together, the
campaign glues together learning methods with innovative physics-based
methods, with iterative algorithms allowing both upstream and downstream
feedback to overcome fundamental limitations of classical \textit{in silico}
drug design~\cite{lee2020scalable-short}. It includes high-throughput
structure-based protein-ligand docking simulations, followed by iterative
refinements to these virtual screening results to filter out compounds that
``show promise'' in biochemical or whole-cell assays, safety and toxicology
tests.

\begin{figure}
  \centering
  \includegraphics[width=0.8\textwidth]{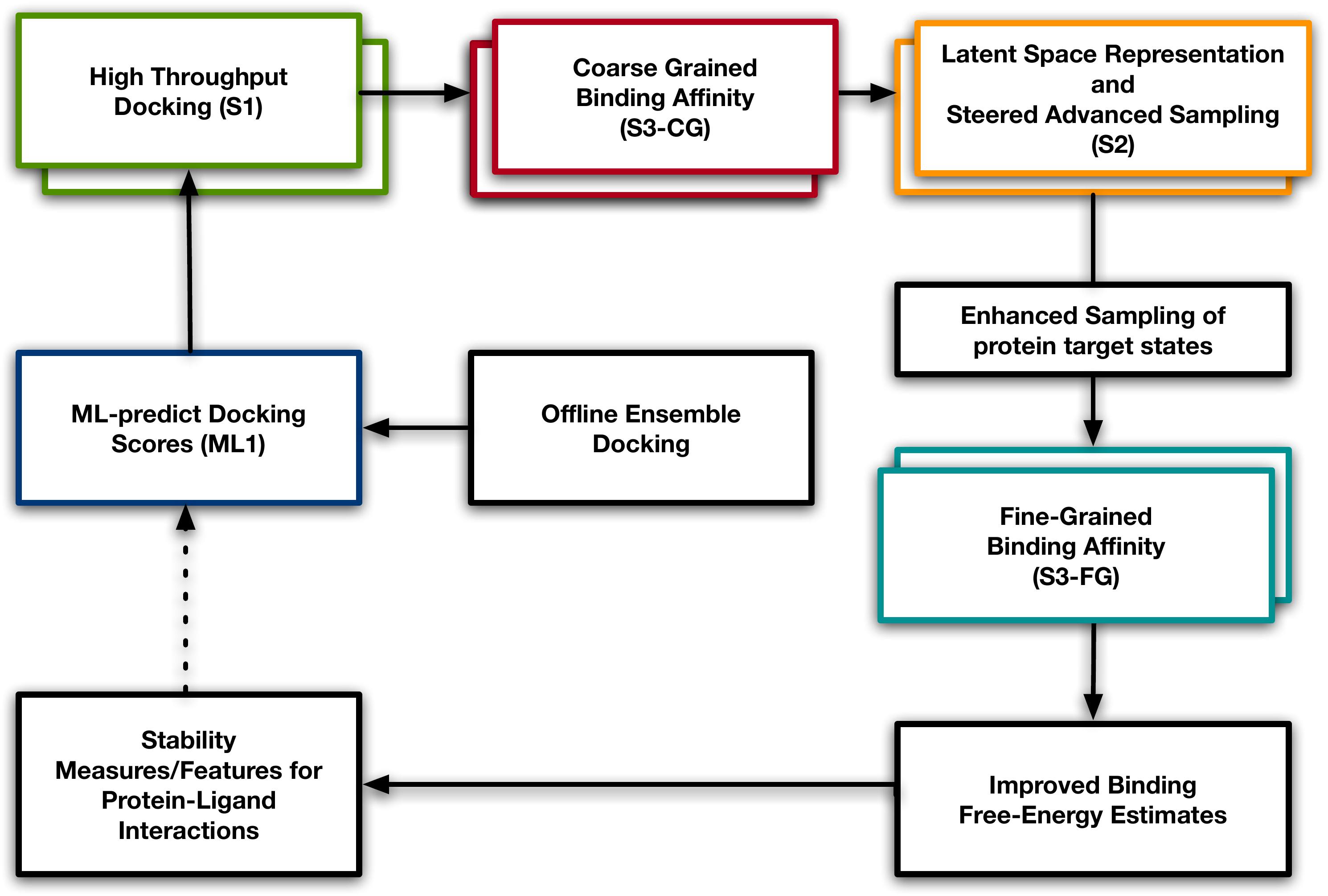}\
  \caption{IMPECCABLE is a virtual drug discovery pipeline, from hit to lead
  through to lead optimization. The constituent components are deep-learning
  based surrogate model for docking (ML1), Autodock-GPU (S1), coarse and
  fine-grained binding free energies (S3-CG and S3-FG) and ML-enhanced MD
  simulations.}\label{fig:impeccablesoln}
\end{figure}

ML techniques overcome the limitations of S1 and S3 by predicting the likelihood
of binding between small molecules and a protein target (ML1), and accelerating
the sampling of conformational landscapes to bound the binding free-energy
values for a given protein-ligand complex (S2). Interfacing ML approaches with
physics-based models (docking and MD simulations), we achieve at least three
orders of magnitude improvement in the size of compound libraries that can be
screened with traditional approaches, while simultaneously providing access to
binding free-energy calculations that can impose better confidence intervals in
the ligands selected for further (experimental or computational) optimization.

S1 is an example of ML-in-HPC mode(i.e., training and using a surrogate in
lieu of computations), while S2~\cite{brace2021achieving} represents a common
instance of the ML-about-HPC mode. Although ML-out-HPC was not implemented on
HPC platforms at the time of publication, prototypes were used to determine
optimal allocation of computational resources across
S1-S3~\cite{woo2021optimal}.

The impact of the algorithmic, methodological and infrastructural innovations
resulted by measuring both raw throughput---defined as ligands per unit time,
scientific performance---defined as effective ligands sampled per unit time, as
well as the quality of ligands
selected~\cite{clyde2021high-short,saadi2020impeccable-short,lee2020scalable-short,babuji2020targeting}.
Thus, the IMPECCABLE~\cite{clyde2021high-short,saadi2020impeccable-short} drug discovery
pipeline is the quintessential example of the learning everywhere paradigm.

\section{Machine Learning and Scientific Workflow Applications}\label{chr-rct}

The coupling of AI/ML methods to HPC simulations, poses unprecedented
challenges to the development of middleware systems to support the execution
of scientific workflow applications on increasingly heterogeneous computing
platforms. We outline three main challenges, and discuss how the six use cases
introduced in \S\ref{chr-lee} address those challenges.


\subsection{Challenges}\label{ssec:challenges}



Traditionally, scientific workflows were defined as either High Throughput
Computing (HTC) or High Performance Computing (HPC). The former came to define
the distributed workflows of the grid era; the later epitomized by complex and
large DAGs of processing. The increasing importance and popularity of
ensembles of HPC simulations, resulted in a convergence of these two primary
modes -- high-throughput of high-performance computing (HT-HPC). And
ultimately workflows involving dependencies between large number of (parallel)
tasks, and represented by DAG task-graph. The current workflow middleware
reflects these dominant paradigms and trends. Moving forward, they will be
supplanted by middleware systems which support ML coupling to HT-HPC workflows
at multiple levels of the application.

Integrating ML methods with HPC simulations, results in three primary classes
of workflows: (1) Hybrid HPC-HTC workflows; (2) ML-coupled workflows, discussed
in
\S\ref{chr-lee}; and (3) Edge-to-center workflows, which typically involve
integrating distributed ML with HPC workflows (e.g., with ML on the edge).
This is rapidly becoming an increasingly important type of workflow with
distributed data production and ML execution, and their need to couple to
large data-centers.



Unsurprisingly, coupling ML to HPC simulations also introduces many challenges
\textemdash{} at application, middleware and resource levels. In this chapter
we focus on three main middleware development challenges related to resource
and task execution management to realize the full potential of ML for
scientific workflow applications: (1) task heterogeneity; (2) adaptive
execution; and (3) application performance.

ML introduces multiple levels of task heterogeneity. Alongside traditional
CPU, GPU and, possibly, multi-node MPI tasks, ML usually requires the
execution of high-throughput function calls, often implemented in an
interpreted language as Python, and that may depend on datasets distributed
across repositories managed by diverse organization and platforms. As a
consequence, the middleware that manages the execution of the workflow
application, has to be able to concurrently schedule, place and execute MPI
executables alongside Python functions with wildly varying execution
lifetimes\textemdash{} the former for hours, the latter for as little as
fractions of seconds.


One of the main scientific reasons to use ML in workflow applications is to
improve the analysis that can be done on the data produced by part of the
tasks of the workflow application. While some tasks progress, ML models can be
used to learn relevant features and better drive the progress of the workflow
at runtime. In order to leverage the potential of ML-based analysis, the
workflow application has to become adaptive, i.e., being able to integrate the
results of ML inferences and alter the workflow graph accordingly, and define
the amount of learning to perform at runtime, especially when that amount
cannot be known in advance, before execution~\cite{brace2021achieving}:
simulations must be paused and restarted with new starting points, and/or a
diverse number, type or size of simulations must be started to account for
changed requirements, based on ML inferences. Further, ML training can vary at
runtime, both in amount per model and across multiple models, when used. That
has consequences for the capabilities of the workflow execution middleware.
Alongside the capability of traversing an acyclic direct graph (DAG) to
produce a concrete execution plan, workflow middleware has to update that DAG,
pausing/restarting the execution of some of its nodes, adding/removing some
nodes, and/or dynamically changing the amount of resources allocated to those
nodes.

Finally, for ML to be useful it must enable improvements in both scientific and
execution performance. On one side, the use of ML modeling and inference needs
to improve the scientific computation that it drives, e.g., the accuracy and/or physically simulated duration. 
On the other side, ML has to effectively and efficiently use available
resources when integrated within a workflow application. Resource efficiency
depends on both the amount of time those resources are used in order to
achieve the planned goal of the workflow application, and the percentage of
available resources utilized to achieve that goal. This means that the
workflow execution middleware has to manage the concurrent execution of
heterogeneous tasks in a way that maximizes resource utilization while
minimizing the workflow application total time to completion.

\subsection{Framework and Middleware Solutions}\label{ssec:middleware}



The ML-enabled workflow frameworks described earlier address some or all the
challenges of task heterogeneity, adaptive execution and framework's
performance (as opposed to scientific performance), at different levels of the
middleware software stack.

Proxima~\cite{zamora2021proxima} is implemented as a Python library used to
wrap a Python function. Based on its inputs, Proxima calculates when to infer
via a surrogate model or running the wrapped function. Inferring via a
surrogate model is often faster than executing the wrapped function, resulting
in an overall speedup. Proxima continually monitors the function execution,
dynamically adapting the surrogate configuration parameters and determining
when to retrain the surrogate model at runtime. While Proxima executes
different types of functions (inference, monitoring, evaluation, configuration
and retraining), it is not optimized for HPC and does not concurrently execute
those different functions at scale. Proxima implements adaptivity, by
retraining at runtime and parametrizing the model. Finally, Proxima
performance as Python library is evaluated in terms of Proxima logic, model
(re)training, surrogate usage, and inference.

Colmena~\cite{ward2021colmena} a general-purpose Python library for steering
ensembles of experiments on HPC computing systems. Colmena is designed to
execute different types of tasks, including: simulation, inference (via
surrogate models) model training, and candidate generation. Different from
Proxima, Colmena is designed to scale on HPC platforms, addressing the
heterogeneity challenge by coordinating the (possibly concurrent) execution of
different types of tasks. Similar to Proxima, Colmena enables adaptivity via surrogate
parameterization and (re)training. Colmena uses Parsl as its runtime, avoiding
a reimplementation  of a ML-specific and general-purpose runtime capabilities. Colmena's
performance is measured in terms of communication overheads (e.g., requests or
result object, and data input or output), and scaling performance with different
task duration, result size, and number of workers.

EXARL~\cite{exarl} is a Python framework build on OpenAI Gym to enable the
implementation of arbitrary reinforcement learning (RL) algorithms and their
execution at scale. EXARL implements agents, each based on a learner/actors
architecture in which each agent concurrently uses a scalable number of
learners. Leaners can be implemented as multi-process or MPI executables; 
multiple agents can be executed concurrently. EXARL does not offer specific
capabilities for mapping and launching its agents, relying on third party
tools like, for example, batch system and an MPI infrastructure. As such,
EXARL does not support task heterogeneity and implementing adaptivity requires
coding capabilities on top of its agents. Performance is currently under
evaluation, in terms of scalability of the size of each learners, and number
of concurrent learners and agents.

MUMMI~\cite{mummi1-short,mummi2-short} is a Python workflow manager that
coordinated the execution of massively parallel multiscale simulations. MUMMI
allows to coordinate the concurrent execution of macro- and micro-scale
simulation tasks, coupling them via ML methods to decide what space of the
macro-scale simulations should be explored by the micro-scale ones. MUMMI uses
the Flux job scheduler to coordinate the scheduling and execution of
heterogeneous tasks on both CPUs and GPUs, and the Maestro workflow plugin to
interface its workflow manager component to Flux. MUMMI enables adaptivity,
allowing (re)training of ML models at runtime and using them for steering the
simulations. MUMMI's performance is evaluated in terms of resource utilization
and number of concurrent simulations executed.

IMPECCABLE~\cite{saadi2020impeccable-short,lee2020scalable-short} is a drug
discovery pipeline that executes heterogeneous tasks (i.e., MD simulations, ML
training and inference) on both CPUs and GPUs at scale. Implemented using
RADICAL-Cybertools as workflow middleware and runtime systems, it also uses
DeepDriveMD. IMPECCABLE enables adaptivity by clustering MD trajectories to
steer the ensemble of MD simulations. This may include either starting new
simulations (i.e., expanding the pool of initial MD simulations), or killing
unproductive MD simulations (i.e., simulations stuck in meta-stable states).
IMPECCABLE also supports runtime evaluation of training of docking
surrogate(s). IMPECCABLE's performance is evaluated in terms of resource
utilization, framework's overheads, and total time to completion of the
pipeline and each of its stages.

Importantly, the capabilities offered by DeepDriveMD and RADICAL-Cybertools are
portable across use cases and computational campaigns. 
DeepDriveMD and RADICAL-Cybertools capabilities which are utilized for
IMPECCABLE, also allowed for coordinating the diverse simulations coupled to
ML models, and automate their execution at scale for the
\#COVIDIsAirborne~\cite{dommer2021covidisairborne} campaign. Work is underway to use RADICAL-Cybertools to support workflow orchestration,
heterogeneous task execution and adaptivity at scale.

\section{Discussion}\label{chr-conclusions}

The success of ML-enabled HPC workflows brings to the forth several challenges
and opportunities: (1) Engineering middleware and frameworks to support for
ML-enabled HPC workflows; (2) ML-HPC Benchmarks to measure both execution and
effective performance; (3) Online ML model engineering, to name just a few.

Consistent with the current workflow application landscape, many ML methods are
being implemented as single-point software solutions, supporting specific
user-facing interfaces, use cases and HPC platforms. Nonetheless, as seen in
\S\ref{ssec:middleware}, some solutions are built over existing middleware,
seeking benefits of well-engineered and general-purpose systems. Thus, one of
the main requirements in middleware engineering for ML and HPC will be to
progressively separate the applications, framework, middleware and platform
concerns, enabling ML support across the stack, without having to code a
plethora of independent solutions that all implement similar capabilities.

Another of the main items of the ML-enabled HPC workflow applications roadmap,
is to promote the integration among existing middleware solutions to support the
development of domain-specific ML frameworks. While the middleware layer should
be domain-agnostic, offering general-purpose resource and runtime management
capabilities, often domain scientists require frameworks tailored to their
programming models and abstractions. For example, some scientists may prefer a
configuration-based interface to set their applications' parameters, while
others require an API to manage parallelism at loops level. The goal will be to
develop frameworks tailored to ML-enabled workflow applications, that
leverage runtime capabilities already available, and expose dedicated
abstractions to the users while hiding low-level details.


Currently, filesystem performance and implementation of in-memory data sharing
are among the main limitations faced by ML-driven workflow applications on HPC
platforms. Often, filesystems become bottlenecks for the I/O intensive
operations required by ML, especially when performed on data continuously
generated at runtime. In-memory approach to data exchange among diverse types
of workflow tasks still requires using task-level
capabilities\cite{choi2019co}. That creates friction between using middleware
capabilities to implement coordination protocols and the need to implement
those protocols within the tasks themselves because of in-memory communication
requirements. That impedes a clean separation of concerns between middleware
and tasks, hindering the development of general-purpose, production-grade
solutions.

Finally, with the growing number of datasets stored on cloud platforms and the
need to leverage diverse programming and computation paradigms, integrating
cloud and HPC resources has become a priority. Developing robust and reliable
solutions for such integrations is a sociotechnical challenge.
Socially, cloud and HPC resources leverage different economic models, making
difficult to reconcile two different resource allocation processes. Technically,
the HPC multitenant batch systems with their non-elastic resource allocations,
heavily biased towards large and long single MPI jobs, does not match the
platform, container and function as a service models implemented by cloud
providers. It will be important to develop resource brokering systems, designed
to seamlessly execute large-scale, ML-enabled workflow applications on diverse
and heterogeneous resources.

Performance will be critical for the future development of ML-enabled workflow
applications. Steady-state performance and resource utilization for
large-scale worfklows is a known challenge. For example, workflows that helped
advance research and response to COVID19 and underpinned the Gordon Bell 2020
Special Prize for COVID19 finalists had impressive peak performance, but
modest steady-state performance. With increasing heterogeneity and temporal
variation in the duration of tasks and services -- as can be expected with
ML-coupled HPC workflows, improving steady-state performance and resource
utilization becomes challenging.

Effective performance, and its measure of scientific improvement
over other methods, will have to be complemented by runtime performance to
assure effective and efficient utilization of available computing resources. In
that context, Benchmarks will play a fundamental role to drive both software and
platform development. Benchmarks will have to be accessible and recognized by
relevant scientific communities, enabling to compare performance among
algorithmic methods and application execution. Without those benchmarks, it will
not be possible to converge towards effective algorithmic solutions and,
importantly, asses how efficiently future platform architectures will support
ML-enabled workflows.

The learning everywhere examples show how ML methods will have to be
integrated at multiple levels within workflow applications and the middleware
that enable their executions. Whereas the real impact will arise from computational campaigns that integrate ML with HPC, ML methods will also play a fundamental role for
the middleware, improving online monitoring, tracing and profiling. 
ML methods will also enable improved scheduling algorithms, essential for the
effective placement of tasks at the upcoming exascale, and preemptive data
staging and caching.

\subsubsection*{Acknowledgements} The authors would like to thank Jack Well
and Tom Gibbs (NVIDIA), and Addi Malviya Thakur (ORNL) for valuable suggestions
on early drafts. SJ acknowledges Geoffrey Fox for many useful discussions. SJ
acknowledges funding from DOE (ECP CANDLE and ExaWorks, and DE-SC0021352), as
well as NSF-1931512 (RADICAL-Cybertools).


\bibliographystyle{unsrt}
\bibliography{main,radical,radical-related,omni,impeccable-references,ws-book-sample}      

\end{document}